\documentclass[english,aps,prl,twocolumn,amsmath,amssymb,showpacs,superscriptaddress,notitlepage,longbibliography]{revtex4-2}
\usepackage{bm}
\usepackage[pdftex]{graphicx,hyperref}
\hypersetup{colorlinks = true, urlcolor = blue, linkcolor = blue, citecolor = blue}
\usepackage{mathtools}
\usepackage{ulem}
\usepackage{color}
\usepackage{float} 
\begin{document}

\title{Inter-orbital spin-triplet superconductivity from altermagnetic fluctuations}

\author{Chen Lu}
\email{luchen@hznu.edu.cn}
\thanks{These two authors contributed equally to this work.}
\affiliation{School of Physics and Hangzhou Key Laboratory of Quantum Matter, Hangzhou Normal University, Hangzhou 311121, China}

\author{Chuang Li}
\thanks{These two authors contributed equally to this work.}
\affiliation{Center for Correlated Matter and School of Physics, Zhejiang University, Hangzhou 310058, China}

\author{Chao Cao}
\email{ccao@zju.edu.cn}
\affiliation{Center for Correlated Matter and School of Physics, Zhejiang University, Hangzhou 310058, China}

\author{Huiqiu Yuan}
\affiliation{Center for Correlated Matter and School of Physics, Zhejiang University, Hangzhou 310058, China}

\author{Fu-Chun Zhang}
\affiliation{Kavli Institute for Theoretical Sciences, University of Chinese Academy of Sciences, Beijing 100190, China}

\author{Lun-Hui Hu}
\email{lunhui@zju.edu.cn}
\affiliation{Center for Correlated Matter and School of Physics, Zhejiang University, Hangzhou 310058, China}

\begin{abstract}
Altermagnetic (AM) fluctuations are a new class of collinear spin fluctuations whose role in mediating superconductivity faces a fundamental tension: their $\Gamma$-point peak favors intra-orbital spin-triplet pairing, while their spin compensation favors inter-orbital singlets. Here, we demonstrate that inversion-symmetry-broken AM fluctuations generically resolve this competition in favor of spin-triplet pairing. As a proof of concept, we study a minimal two-orbital model with two van Hove singularities. The broken inversion symmetry induces momentum-orbital locking—the same orbital dominates at opposite momenta—enhancing the triplet channel. Crucially, a subdominant fluctuation channel arising from inter-van-Hove nesting provides an internal Josephson coupling that locks the phase difference between triplet pairs on different orbitals. We find this coupling changes sign ($+$ to $-$) upon a crossover from AM-dominant to ferromagnetic-dominant fluctuations. The resulting $\pi$-phase difference manifests as a $\tau_z$-type order parameter, $c_{\bm{k},1\uparrow}c_{-\bm{k},1\uparrow} - c_{\bm{k},2\uparrow}c_{-\bm{k},2\uparrow}$. Although intra-orbital in the original basis, its orbital-nontrivial character, as manifested by its equivalence to inter-orbital pairing under rotation, defines a general \textit{inter-orbital spin-triplet superconductivity}. This state is distinct from the $\tau_0$-triplet pairing mediated by ferromagnetic fluctuations, as evidenced by the canceled intra-orbital supercurrent in a Josephson junction between them.
\end{abstract}

\maketitle

\paragraph{{\color{blue}Introduction}} 
In unconventional superconductors, the pairing mechanism arises not solely from electron–phonon coupling, but primarily from electron–electron interactions~\cite{Sigrist1991rmp,Dagotto1994rmp,Harlingen1995rmp,mineev1999introduction,joynt2002superconducting,anderson2004jpcm,Balatsky2006rmp,davis2013concepts,stewart2017unconventional,smidman2017ropp,Smidman2023rmp}.
The repulsive interactions can become effectively attractive through the mediation of spin fluctuations, giving rise to the well-established theory of fluctuation-mediated superconductivity~\cite{Berk1966prl,Leggett1975rmp,Bickers1989prl,schrieffer1989dynamic,moriya1990antiferromagnetic,moriya2000spin,scalapino2012common}.
This framework has been successfully applied to various experimentally observed unconventional superconducting materials. A prominent example is found in iron-based superconductors, where N\'eel antiferromagnetic (AFM) fluctuations are believed to drive an extended $s_{\pm}$-wave pairing state with a sign change between the $\Gamma$ and $M$ points~\cite{chubukov2008magnetism,paglione2010high,stewart2011superconductivity,wang2011electron,dai2012natphys,Dagotto2013rmp,chen2014nsr,dai2015antiferromagnetic,fernandes2016ropp,fernandes2022nature}.
The ferromagnetic (FM) fluctuations can mediate spin-triplet superconductivity, with candidate materials including K$_2$Cr$_3$As$_3$~\cite{Bao2015prx,zhou2017theory,wu2015triplet,Chen2018ropp,yang2021spin}, UTe$_2$~\cite{ran2019science,jiao2020nature,hayes2021multicomponent,aoki2022unconventional,lewin2023ropp}, and CeSb$_2$~\cite{zhang2022kondo,squire2023superconductivity,shan2025emergent}.

Spin-triplet superconductors are of particular interest due to their non-trivial topological properties and potential for hosting Majorana quasiparticles, which are crucial for topological quantum computing~\cite{Read2000prb,kitaev2003fault,Nayak2008rmp,qi2011rmp,beenakker2013search,sarma2015majorana,ando2015arcmp,sato2017topological,flensberg2021engineered}.
This immense potential fuels the search for new triplet-pairing mechanisms and material realizations.
However, the exploration has largely remained within the traditional dichotomy of FM fluctuations. A fundamental open question is whether other classes of magnetic fluctuations can also generate robust spin-triplet superconductivity.

Here, we address this question by considering dominant altermagnetic (AM) fluctuations.
Altermagnetism is a recently identified third class of collinear magnetic orders~\cite{naka2019spin,ahn2019antiferromagnetism,hayami2019momentum,vsmejkal2020crystal,yuan2020giant,shao2021spin,mazin2021prediction,ma2021multifunctional,yuan2021prm,vsmejkal2022beyond,vsmejkal2022emerging,bai2024altermagnetism,jungwirth2024altermagnets,fender2025altermagnetism,song2025altermagnets}, and has been experimentally observed~\cite{fedchenko2024ruo2,lin2024ruo2,gonzalez2023prl,krempasky2024MnTe,lee2024MnTe,osumi2024MnTe,reichlova2024observation,liu2024chiral,reimers2024CrSb,ding2024CrSb,yang2025three,jiang2025metallic,zhang2025crystal}.
AM fluctuation is a less-explored type of collinear spin fluctuation capable of mediating unconventional superconductivity. Their defining traits, however, present a dilemma: the $\bm{Q}=\bm{0}$ propagation vector (like FM) promotes intra-sublattice spin-triplet pairing~\cite{wu2025intra,ma2025possible}, while its spin compensation (like N\'eel AFM) favors a competing inter-sublattice singlet channel~\cite{wu2025intra}. 
The mechanism for stabilizing spin-triplet pairing remains unclear.

Our work reveals that inversion-symmetry-broken AM fluctuations generically favor spin-triplet pairing. As a proof of concept, we study a minimal two-orbital system with Fermi energy near two van Hove singularities (VHS). The intra-VHS nesting produces dominant $\bm{Q}=0$ AM fluctuations, while momentum–orbital locking enhances intra-orbital triplet pairing. Crucially, the inter-VHS nesting generates a subdominant fluctuation channel that serves as an internal $\pi$-phase Josephson coupling, which locks the phase between spin-triplet pairs and yields an orbital-nontrivial $\tau_z$-type order parameter. A unitary rotation (e.g., $\tau_{z}\leftrightarrow\tau_{x}$) reveals this intra-orbital state to be fundamentally an \textit{inter-orbital spin-triplet superconductor}, distinct from the FM-fluctuation-mediated $\tau_0$-triplet. This distinction is directly testable via a canceled intra-orbital supercurrent in a $\tau_z$- and $\tau_0$-triplet Josephson junction.

\begin{figure}[t]
\centering
\includegraphics[width=0.48\textwidth]{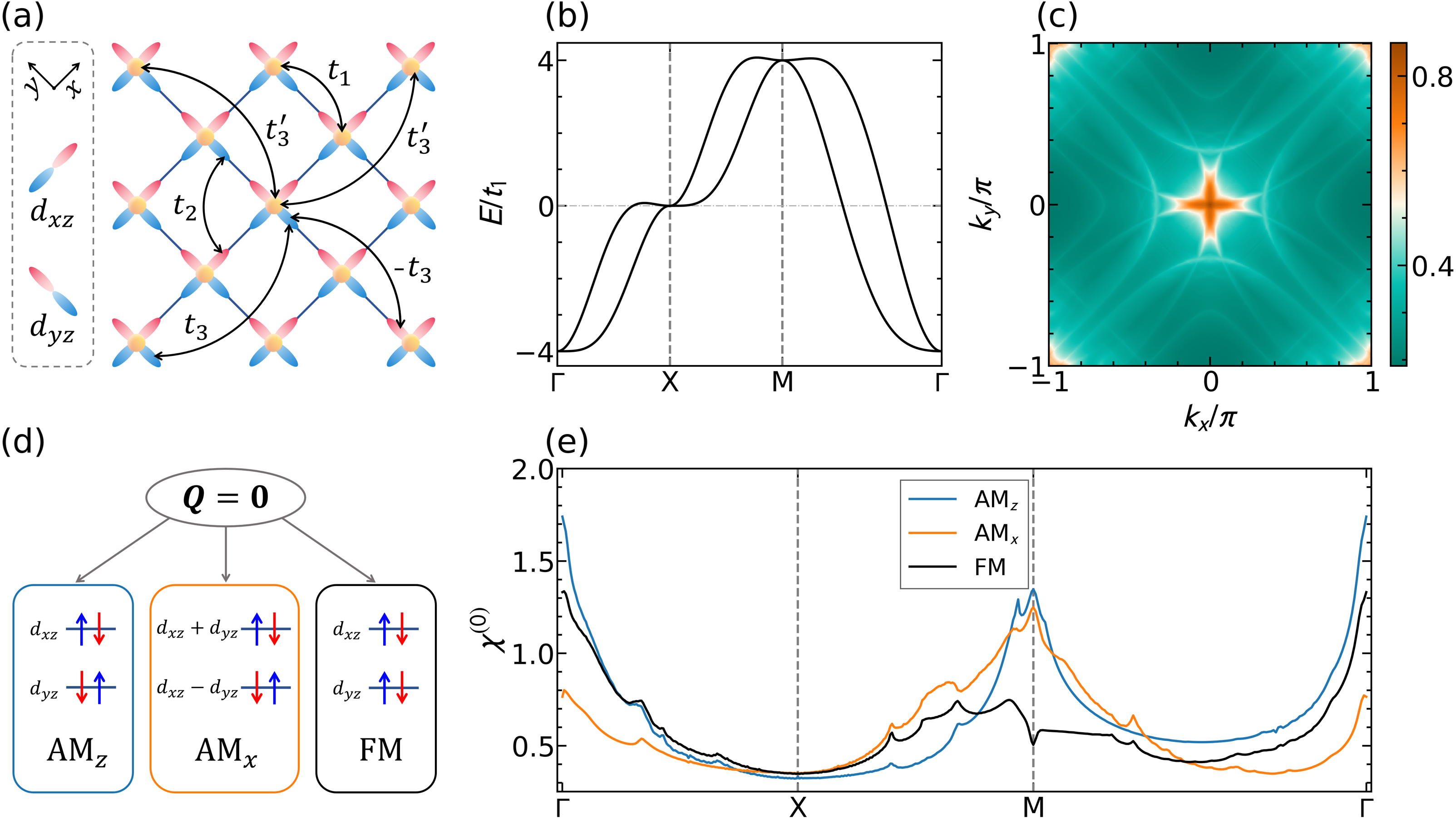}
\caption{Band structure and dominant magnetic fluctuations. 
(a) Schematic of the two-orbital square lattice model.
(b) Band structure along high-symmetry paths, with the Fermi level (gray dashed line) tuned near van Hove singularities.
(c) Momentum distribution of the largest eigenvalue of the static bare susceptibility $[\chi^{(0)}(\bm{k},i\omega= 0)]^{l_1 l_1}_{l_2 l_2}$, showing a pronounced peak at the $\Gamma$ point.
(d) Three possible $\bm{Q}=\bm{0}$ onsite magnetic orders: $\tau_z$- and $\tau_x$-type altermagnetism (AM), and ferromagnetism (FM). Blue ($\color{blue}\uparrow$) and red ($\color{red}\downarrow$) arrows represent the spin polarization of local moments.
(e) Momentum dependence of the bare susceptibility for the AM$_z$, AM$_x$, and FM channels, revealing the AM$_z$ fluctuation as the dominant fluctuation.
}
\label{fig1}
\end{figure}

\paragraph{{\color{blue}Altermagnetic fluctuations}}
We begin by defining dominant AM fluctuations based on the standard spin-spin correlation functions. Here, we consider a minimal two-orbital system (e.g.,~atomic, sublattice, or layer). Specifically, we study a tight-binding model on a two-dimensional square lattice, where each site hosts two atomic orbitals ($d_{xz}$, $d_{yz}$) [Fig.~\ref{fig1}(a)]. The normal-state Hamiltonian reads
\begin{align}\label{Hamiltonian_tb}
{\cal H}_{0}(\bm{k}) &= \varepsilon_0(\bm{k}) \tau_0 + \varepsilon_1(\bm{k}) \tau_x + \varepsilon_2(\bm{k}) \tau_z,
\end{align}
where $\varepsilon_0(\bm{k}) = -2t_1[\cos(k_x)+\cos(k_y)] - 2t^{\prime}_3[\cos(2k_x)+\cos(2k_y)]-\mu$ with $\mu$ the chemical potential, $\varepsilon_1(\bm{k})=4t_2\sin(k_x)\sin(k_y)$, $\varepsilon_2(\bm{k})=2t_3[\cos(2k_x)-\cos(2k_y)]$, and $\tau_{x,y,z}$ represent the Pauli matrices acting on the orbital space. Here, $t_1$ (1st-neighbor) and $t^{\prime}_3$ (3rd-neighbor) are orbital-independent hoppings, while $t_2$ (2nd-neighbor) and $t_3$ (3rd-neighbor) are orbital-dependent, acting as hybridizations between the two orbitals. We use the parameter set $t_1=1.0$, $t_2=0.51$, $t_3=0.29$, and $t_3^{\prime}=0.04$. The resulting band structure is shown in Fig.~\ref{fig1}(b). The Fermi level, set at $\mu=-0.16$ (gray dashed line), lies near van Hove singularities (VHS), which can promote an instability toward $\bm{Q}=\bm{0}$ magnetic orders~\cite{yu2025altermagnetism}.

The non-interacting spin fluctuations are described by the bare spin susceptibility, a fourth-rank tensor in the orbital space defined as,
\begin{align}\label{chi0}
\begin{split}
&[\chi^{(0)}(\bm{k},i \omega)]^{l_1l_2}_{l_3l_4} \equiv 
\frac{1}{N} \sum_{\bm{k}'} \sum_{\alpha,\beta}  [\xi^{\alpha}_{l_1}(\bm{k}')]^\ast \xi^{\beta}_{l_2}(\bm{k}'+\bm{k}) \times \\
&\qquad\quad [\xi^{\beta}_{l_3}(\bm{k}'+\bm{k})]^\ast  \xi^{\alpha}_{l_4}(\bm{k}')\frac{\eta_F(\varepsilon^{\beta}_{\bm{k}'+\bm{k}})-\eta_F(\varepsilon^{\alpha}_{\bm{k}'})}{i\omega+\varepsilon^{\alpha}_{\bm{k}'}- \varepsilon^{\beta}_{\bm{k}'+\bm{k}}}, 
\end{split}
\end{align}
where $l_1, l_2, l_3, l_4$ are orbital indices, $\alpha, \beta$ are band indices, $N$ is the number of lattice sites, $\varepsilon^{\alpha}_{\bm{k}}$ and $\xi^{\alpha}(\bm{k})$ are the $\alpha$-th eigenvalue and eigenvector of ${\cal H}_{0}(\bm{k})$, respectively, and $\eta _F$ is the Fermi-Dirac distribution function. 
Considering the diagonal elements in orbital space ($l_1=l_2$, $l_3=l_4$) reduces the $\chi^{(0)}$-tensor to a matrix, which corresponds to the spin-spin correlation function~\footnote{This matrix $[\chi^{(0)}(\bm{k})]_{l_3l_3}^{l_1l_1}$ is equivalent to: $[\chi_{zz}^{(0)}(\bm{k})]_{l_3 l_3}^{l_1 l_1}=\frac{1}{N}\sum_{i,j} e^{-i\bm{k}\cdot(\bm{R}_i-\bm{R}_j)}\langle S_{l_1}^z(\bm{R}_i) S_{l_3}^z(\bm{R}_j) \rangle_0$.}.
The largest eigenvalue of this matrix exhibits a pronounced peak at the $\Gamma$ point [Fig.~\ref{fig1}(c)], indicating a dominant instability toward $\mathbf{Q}=\bm{0}$ magnetic order. This suggests three possible magnetic ordering channels, with their orbital-resolved spin configurations illustrated in Fig.~\ref{fig1}(d). Due to the broken inversion symmetry $\mathcal{P}_{xy}$ (which interchanges the $d_{xz}$ and $d_{yz}$ orbitals) by the $t_3$ term in Eq.~\eqref{Hamiltonian_tb}, these orders are represented in orbital space by the matrices $\tau_z, \tau_x, \tau_0$; the first two describe AM order and the last describes FM order~\footnote{The altermagnetism requires a crystalline symmetry that connects the two magnetic sublattices; in our model, this role is played by the four-fold rotational symmetry.}.
To determine the leading instability or dominant fluctuation among these channels~\cite{Roig2024prb,wang2025spin,lu2025breakdown}, we calculate the static bare susceptibility for each as,
\begin{align}\label{chiO}
\chi^{(0)}_{\alpha}(\bm{k}) = \frac{1}{2} \sum_{l_1l_2l_3l_4}  [\bar{\cal O}_\alpha]_{l_1l_2} [\bar{\cal O}_\alpha]_{l_3l_4} [\chi^{(0)}(\bm{k},0)]^{l_1l_2}_{l_3l_4},
\end{align}
with $\bar{\cal O}_{\text{AM}_z} = \tau_z$, $\bar{\cal O}_{\text{AM}_x} = \tau_x$ and $\bar{\cal O}_{\text{FM}} = \tau_0$. As shown in Fig.~\ref{fig1}(e), the AM$_z$ channel (blue curve) exhibits the strongest peak at the $\Gamma$ point, originating from intra-VHS nesting, which represents the dominant fluctuation in the system. We thus identify these as the dominant \textit{altermagnetic fluctuation} in the system and now explore its role in mediating spin-triplet pairing.

\paragraph{{\color{blue}Inter-orbital spin-triplet pairing}}
The superconducting pairing symmetry mediated by spin fluctuations can be determined within the multi-orbital random phase approximation (RPA) framework~\cite{scalapino1986d,hamann1969properties}. We consider the standard on-site repulsive Hubbard–Hund interaction Hamiltonian,
${\cal H}_{int} = H_U + H_V + H_J$, with $H_U=U \sum _{\bm{i},\tau} n_{\bm{i}\tau \uparrow }n_{\bm{i}\tau \downarrow }$, $H_V=V\sum_{\bm{i}s,s^{\prime}} n_{\bm{i},x,s}n_{\bm{i},y,s^{\prime}}$, and $H_J= \\ J_H \sum _{\bm{i}}  [ \sum _{s,s^{\prime}} c^{\dagger}_{\bm{i},x,s}c^{\dagger}_{\bm{i},y,s^{\prime}}c_{\bm{i},x,s^{\prime}}c_{\bm{i},y,s} + c^{\dagger}_{\bm{i},x,\uparrow} c^{\dagger}_{\bm{i},x,\downarrow}  c_{\bm{i},y,\downarrow} c_{\bm{i},y,\uparrow} \\ + h.c. ]$.
Here, $c_{\bm{i},\tau,s}$ is the electron annihilation operator at site $\bm{i}$ with orbital $\tau$ and spin $s$, $n_{i\tau s}=c_{\bm{i},\tau,s}^\dagger c_{\bm{i},\tau,s}$ is the density operator, $\tau=\{x,y\}$ labels the $\{d_{xz}, d_{yz}\}$ orbitals, and $s=\{\uparrow,\downarrow\}$ denotes spin. The interaction parameters $U$, $V$, $J_H$ represent intra-orbital repulsion, inter-orbital repulsion, and Hund's coupling (including pair hopping), respectively. The rotational symmetry of the orbital space imposes the constraint $U = V + 2J_H$. The RPA-renormalized susceptibility for Eq.~\eqref{chiO} is
\begin{align} \label{chiO_RPA}
\chi_{\alpha}^{\text{RPA}}(\bm{k}) &= \frac{1}{2} \sum_{l_1l_2l_3l_4}  [ \bar{\cal O}_\alpha ]_{l_1l_2} [  \bar{\cal O}_\alpha ]_{l_3l_4} [\chi^{\text{RPA}}_{\text{spin}}(\bm{k})]^{l_1l_2}_{l_3l_4} , 
\end{align}
where the full spin susceptibility $\chi^{\text{RPA}}_{\text{spin}}(\bm{k}) = \chi^{(0)}(\bm{k}) [I - \chi^{(0)}(\bm{k}) {\cal U}_{s}]^{-1}$ is renormalized by interactions. Here, $I$ denotes the identity matrix and ${\cal U}_{s}$ is the interaction matrix in the spin channel, which is proportional to $U$. Hence, we enhance spin fluctuations by increasing $U$ and they become most pronounced when $U\to U_c$. Our analysis focuses on the regime $U < U_c$, where strong fluctuations mediate unconventional superconductivity. Moreover, $J_H$ typically favors FM order: our calculations confirm the AM$_z$ fluctuation dominates at low $J_H$, crossing over to the FM fluctuation at large $J_H$ [see Sec.~A of Supplementary Material (SM)].

\begin{figure}[t]
\centering
\includegraphics[width=0.48\textwidth]{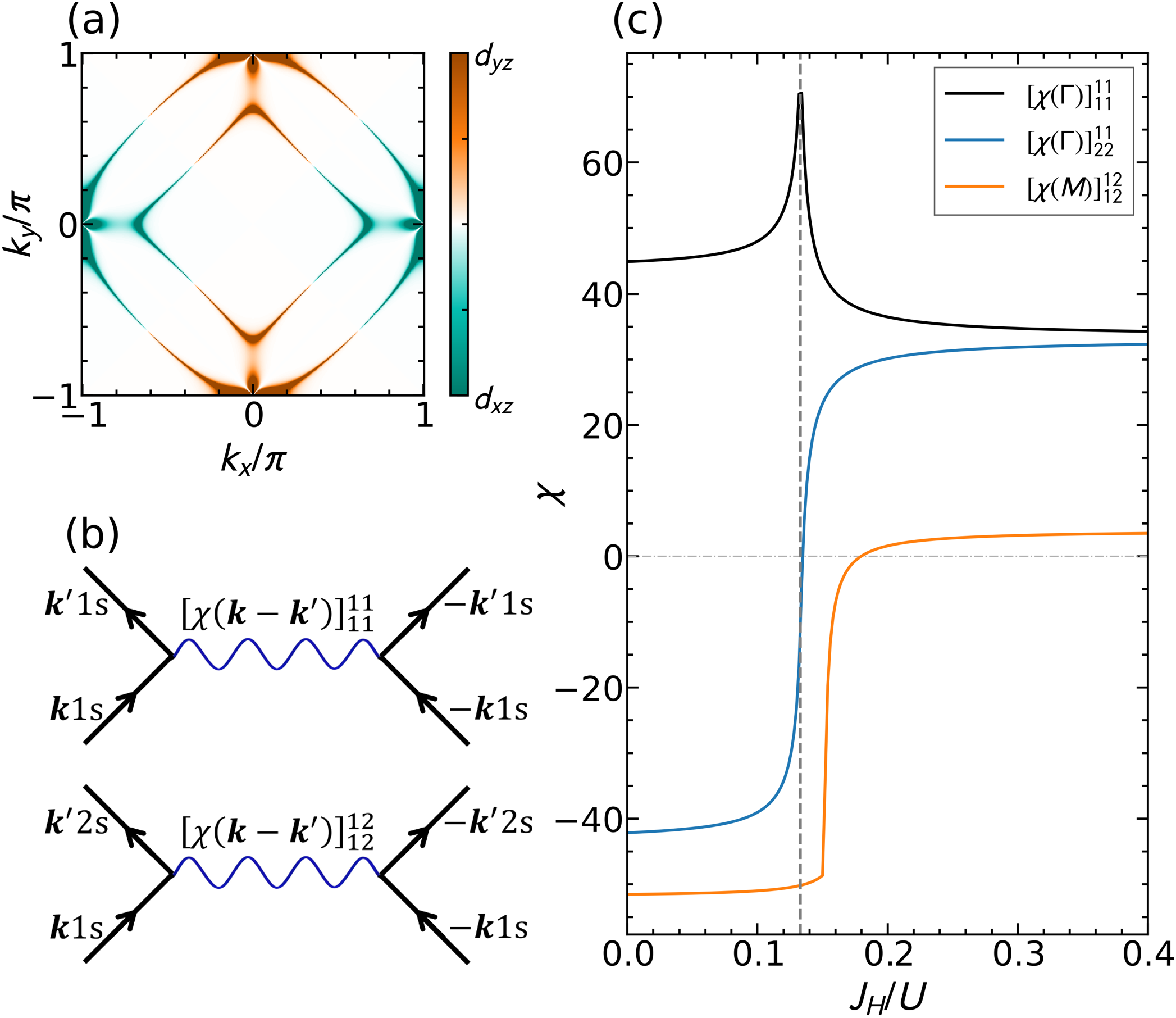}
\caption{Pairing mechanism.
(a) Orbital-polarized Fermi surfaces near van Hove singularities: $d_{xz}$ ($d_{yz}$) character around the X (Y) point.
(b) Feynman diagrams for the two key pairing channels: intra-orbital pairing mediated by $[\chi^{\text{RPA}}_{\text{spin}}(\bm{q})]^{11}_{11}$ and the pair-hopping mediated by $[\chi^{\text{RPA}}_{\text{spin}}(\bm{q})]^{12}_{12}$.
(c) Evolution of the pairing vertices $[\chi^{\text{RPA}}_{\text{spin}}(\Gamma)]^{11}_{11}$ (black), $[\chi^{\text{RPA}}_{\text{spin}}(\Gamma)]^{11}_{22}$ (blue), and $[\chi^{\text{RPA}}_{\text{spin}}(\text{M})]^{12}_{12}$ with $J_H/U$ and $U=0.99U_c$.
}
\label{fig2}
\end{figure}

The attractive pairing interaction primarily originates from the $2\times 2$ block $[\chi^{\text{RPA}}_{\text{spin}}(\bm{k})]^{l_1l_1}_{l_3l_3}$, structured as nearly $\propto \tau_0 \mp \tau_x$ for AM$_z$ and FM channels, respectively~\footnote{Our RPA calculations at $U=0.99U_c$ yield the following results for the $[\chi^{\text{RPA}}_{\text{spin}}(\Gamma)]^{l_1l_1}_{l_3l_3}$ matrix: near the AM$_z$ phase ($J_H/U=0.02$), it is $\begin{pmatrix} 45.1 & -41.9 \\ -41.9 & 45.1 \end{pmatrix}$, while near the FM phase ($J_H/U=0.3$), it becomes $\begin{pmatrix} 34.7 & 31.9 \\ 31.9 & 34.7 \end{pmatrix}$. More details are shown in Fig.~\ref{fig2}(c).}.
The $\tau_0$ component, common in both AM$_z$ and FM fluctuations, promotes intra-orbital spin-triplet pairing; whereas the $\mp\tau_x$ component dictates a distinct inter-orbital channel: spin-singlet for AM$_z$ and spin-triplet for FM fluctuations~\cite{wu2025intra}. However, the triplet channel is inherently favored due to the dominant AM fluctuations. By definition, $\mathcal{P}_{xy}$ symmetry breaking is required, which enforces momentum-orbital locking: $d_{xz}$- and $d_{yz}$-polarized states around $\bm{X}$ and $\bm{Y}$, respectively [Fig.~\ref{fig2}(a)]. This naturally suppresses inter-orbital pairing. Consequently, the positive $[\chi^{\text{RPA}}_{\text{spin}}(\bm{k})]^{11}_{11} \tau_0$ term for $\bm{k}\sim \Gamma$ mandates spin-triplet pairing, via the standard spin-fluctuation exchange: 
\begin{align}
{\cal H}_{\text{pair}}^{(1)} \propto -[\chi^{\text{RPA}}_{\text{spin}}(\bm{k}-\bm{k}')]^{11}_{11} c_{\bm{k},l,s}^\dagger c_{-\bm{k},l,s}^\dagger c_{-\bm{k}',l,s} c_{\bm{k}',l,s}.    
\end{align}
We classify the spin fluctuation as AM-type based on its dominant peak at $\bm{Q}=0$; however, the pairing interaction is mediated by the full momentum structure of these fluctuations, not solely the $\bm{Q}=0$ component. A key secondary process is the pair-hopping term [Fig.~\ref{fig2}(b)],
\begin{align}
{\cal H}_{\text{pair}}^{(2)} \propto -[\chi^{\text{RPA}}_{\text{spin}}(\bm{k}-\bm{k}')]^{12}_{12} c_{\bm{k},l,s}^\dagger c_{-\bm{k},l,s}^\dagger c_{-\bm{k}',\bar{l},s} c_{\bm{k}',\bar{l},s},
\end{align}
which dominates at $\bm{k}-\bm{k}'=(\pi,\pi)\equiv M$. As shown in Fig.~\ref{fig2}(c), $[\chi^{\text{RPA}}_{\text{spin}}(\Gamma)]^{11}_{11}$ (black) remains positive across the phase diagram, whereas $[\chi^{\text{RPA}}_{\text{spin}}(\text{M})]^{12}_{12}$ (orange) changes sign, marking a transition from a $\pi$- to a $0$-phase Josephson coupling. This sign reversal underpins two distinct triplet states: FM fluctuations mediate a $\tau_0$-triplet, while AM$z$ fluctuations generate a $\tau_z$-triplet, characterized by $\langle c_{-\bm{k},x,s} c_{\bm{k},x,s} \rangle = - \langle c_{-\bm{k},y,s} c_{\bm{k},y,s} \rangle $. A rotation to the bonding-antibonding basis transforms the $\tau_z$-triplet into a $\tau_x$-triplet. Thus, the dominant AM fluctuation mediates an \textit{inter-orbital spin-triplet superconductivity}.

\paragraph{{\color{blue}Numerical results}}
We next provide numerical confirmation of the above argument by computing the interaction renormalization from spin fluctuations in the subcritical regime ($U < U_c$). By setting $U = 0.99U_c$, we enhance fluctuations without triggering magnetic order. Since spin–orbital coupling is absent in our system, the resulting effective interaction Hamiltonian is,
\begin{align}\label{Veff1}
V_{\text{eff}} = \frac{1}{N} \sum_{l_1,l_2,l_3,l_4} \sum_{\bm{k}\bm{k}^{\prime}} \Gamma^{l_1l_2}_{l_3l_4}(\bm{k},\bm{k}^{\prime}) {\cal P}_{l_1,l_2}^\dagger(\bm{k}) {\cal P}_{l_3,l_4}(\bm{k}'), 
\end{align}     
where ${\cal P}_{l_1,l_2}^\dagger(\bm{k}) = c^{\dagger}_{\bm{k}, l_1, \uparrow} c^{\dagger}_{-\bm{k}, l_2,\downarrow}$ and ${\cal P}_{l_3,l_4}(\bm{k}') =  c_{-\bm{k}', l_3,\downarrow} \\ c_{\bm{k}', l_4,\uparrow}$ are Cooper pair creation and annihilation operators, and $\Gamma^{l_1l_2}_{l_3l_4} (\bm{k},\bm{k}^{\prime})$ denotes the interaction vertex. As established previously, $V_{\text{eff}}$ contains attractive channels that drive superconducting instability~\cite{kohn1965new}. To distinguish between triplet pairing mediated by AM fluctuations and that driven by FM fluctuations, we vary $J_H$, examining the system near the $\text{AM}_z$ phase ($J_H/U=0.02$) and the $\text{FM}$ phase ($J_H/U=0.3$). The detailed form of the vertex function is provided in Sec.~B of SM.

\begin{figure}[b]
\centering
\makebox[\linewidth][c]{
\includegraphics[width=0.48\textwidth]{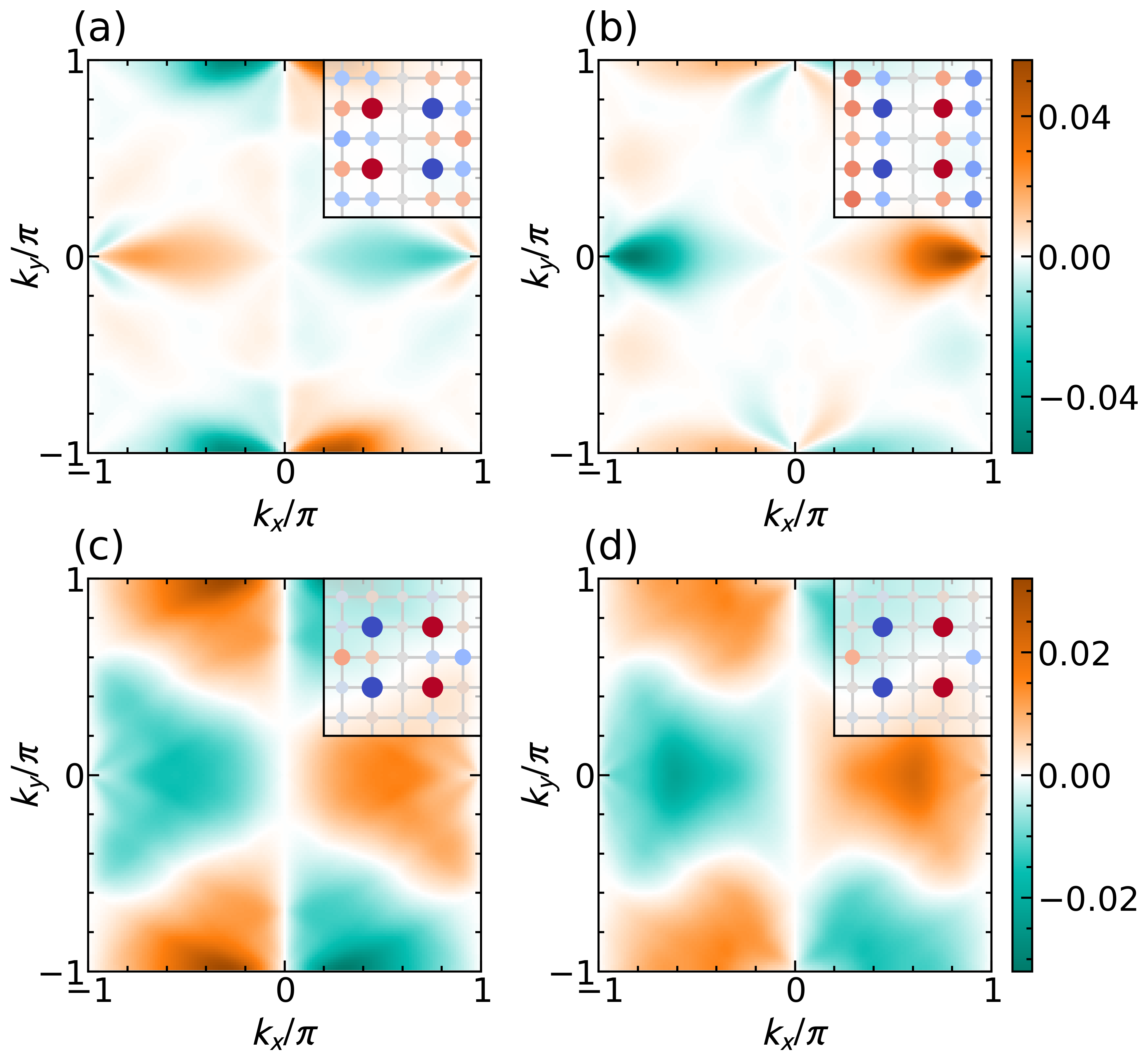}
}
\caption{Fluctuation-pairing correspondences.
The orbital-resolved gap functions, $\Delta^x_{d_{xz},d_{xz}}(\bm{k})$ and $\Delta^x_{d_{yz},d_{yz}}(\bm{k})$, for (a,b) AM$_z$-mediated ($J_H/U=0.02$) and (c,d) FM-mediated ($J_H/U=0.3$) pairings. The phase relation between orbitals differentiates the $\tau_z$-triplet from the $\tau_0$-triplet state, with the corresponding real-space patterns shown in the insets.
}
\label{fig3}
\end{figure}

We then determine the superconducting order parameter by solving ${\cal H}_{\text{tot}}={\cal H}_0+V_{\text{eff}}$ [Eqs.~\eqref{Hamiltonian_tb} and \eqref{Veff1}] self-consistently. The orbital-dependent pairing functions are defined as $\Delta_{l_1,l_2}(\bm{k}) = \langle c_{\bm{k}, l_1, \uparrow} c_{-\bm{k}, l_2, \downarrow} \rangle$.
In Fig.~\ref{fig3}, panels (a-b) and (c-d) display the momentum-space pairing functions for different orbitals induced by AM$_z$ and FM fluctuations, respectively. The four-fold rotational symmetry of our model yields degenerate linear combinations of $p_x$- and $p_y$-like states in the mean-field solution. To analyze the $p_x$-wave component, we define $\Delta_{l_1,l_2}^x(\bm{k}) = \Delta_{l_1,l_2}(k_x,k_y) + \Delta_{l_1,l_2}(k_x,-k_y)$. The odd-parity nature of the spin-triplet pairing, $\Delta_{l,l}^x(k_x,k_y) = - \Delta_{l,l}^x(-k_x,k_y)$, is confirmed for both AM$_z$ [Figs.~\ref{fig3}(a-b)] and FM [Figs.~\ref{fig3}(c-d)] fluctuations. The crucial distinction between these two spin-triplet states lies in the relative phase of the order parameters on $d_{xz}$ and $d_{yz}$ orbitals,
\begin{align} \label{eq-AM-FM-triplet}
\begin{cases}
\text{AM}_z\text{-mediated triplet: } \Delta^x_{d_{xz},d_{xz}} = -\Delta^x_{d_{yz},d_{yz}}, \\  
\text{FM}\text{-mediated triplet: } \Delta^x_{d_{xz},d_{xz}} = \Delta^x_{d_{yz},d_{yz}}.  \\ 
\end{cases}    
\end{align}
This is the central finding of this work. For the AM$_z$-mediated case, the order parameters on $d_{xz}$ and $d_{yz}$ orbitals exhibit a $\pi$-phase difference, in sharp contrast to the in-phase, $\tau_0$-triplet pairing mediated by FM fluctuations. Furthermore, we note that our numerical calculations, which incorporate the full spin fluctuation beyond pure AM or FM channels, yield pairing functions that are mixtures of $\tau_z$- and $\tau_0$-triplet components.

To further characterize the pairing symmetry, we compute the real-space pairing function $\Delta_{l_1, l_2}(\bm{r}) = \langle c_{\bm{i}, l_1, \uparrow} c_{\bm{i}+\bm{r},l_2,\downarrow} \rangle$ via Fourier transformation. With the reference site at $\bm{i}=(0,0)$, the dominant pairing correlations reside on the next-nearest-neighbor bonds, as shown in the insets of Fig.~\ref{fig3}, where the marker color and size scale with the imaginary part of $\Delta_{l_1, l_2}(\bm{r})$. We extract the amplitudes $\tilde{\Delta}_{x,x} \equiv \text{Im}[\Delta_{d_{xz}, d_{xz}}(1,1)]$ and $\tilde{\Delta}_{y,y} \equiv \text{Im}[\Delta_{d_{yz}, d_{yz}}(1,1)]$. The corresponding Bogoliubov-de Gennes (BdG) Hamiltonian for the AM$_z$-mediated spin-triplet superconductivity is,
\begin{align} \label{eq-bdg-hamsc}
{\cal H}_{\text{BdG}}(\bm{k}) = 
\begin{pmatrix}
{\cal H}_0(\bm{k}) & \Delta_t(\bm{k}) \tau_z \\
\Delta_t(\bm{k}) \tau_z & -{\cal H}_0^\ast(\bm{k})
\end{pmatrix},
\end{align}
where $\Delta_t(\bm{k}) = \Delta_0 \sin(k_x)\cos(k_y)$ with $\Delta_0 = 2(\tilde{\Delta}_{x,x}-\tilde{\Delta}_{y,y})$. Replacing $\Delta_t(\bm{k})\tau_z$ with $\Delta_t(\bm{k})\tau_0$ and setting $\Delta_0 = 2(\tilde{\Delta}_{x,x}+\tilde{\Delta}_{y,y})$ yields the FM-mediated case. The simplified model in Eq.~\eqref{eq-bdg-hamsc} exhibits nodal lines along $k_x=0,\pi$ or $k_y=\pm\pi/2$, which slightly deviate from the RPA results in Fig.~\ref{fig3}, due to our retention of only the dominant channels [Fig.~\ref{fig2}(b)]. Furthermore, by tuning parameters to enhance AM$_x$ fluctuations, we consistently obtain $\tau_x$-triplet pairing, as detailed in Sec.~C of SM.

\begin{figure}[t]
\centering
\includegraphics[width=0.48\textwidth]{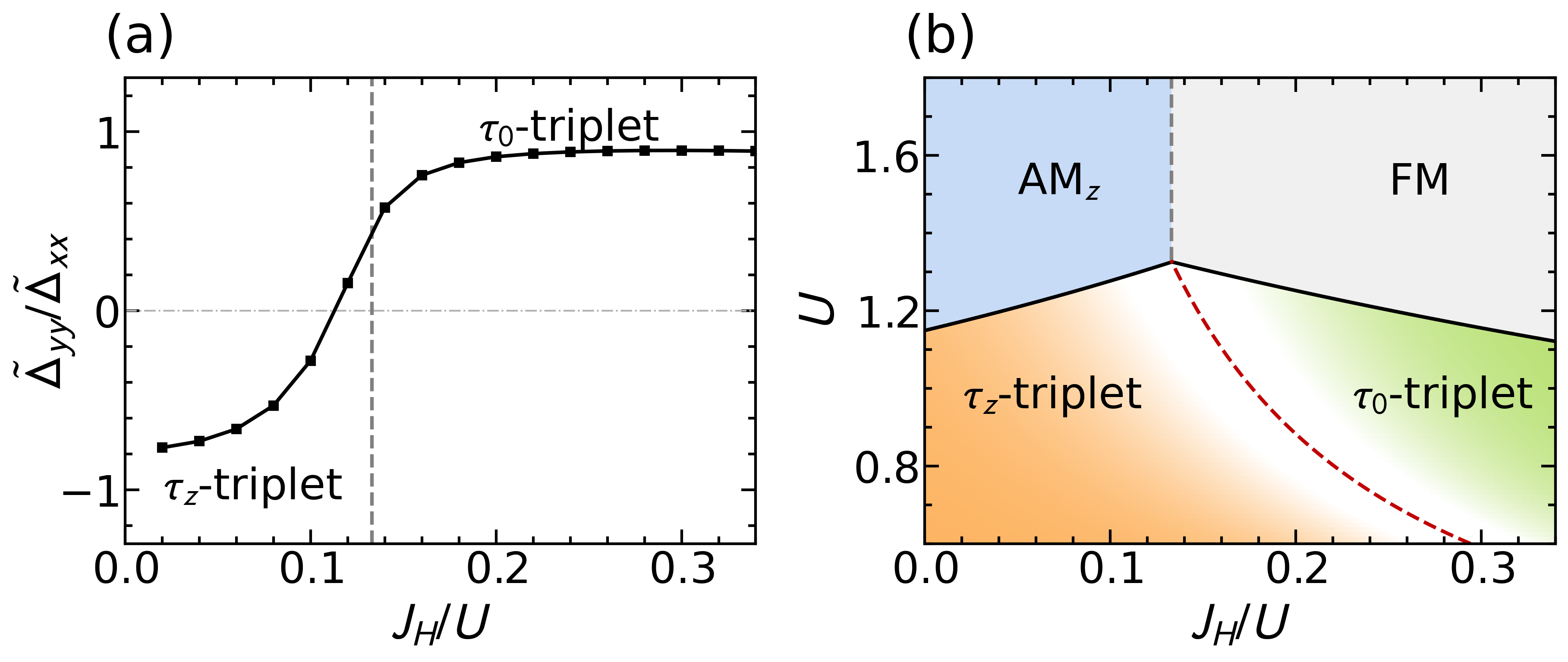}
\caption{Phase diagram.
(a) Evolution of the pairing amplitude ratio $\tilde{\Delta}{xx}/\tilde{\Delta}{yy}$ with $J_H/U$ at $U=0.99U_c$, signaling the transition between $\tau_z$- and $\tau_0$-triplet states.
(b) Full $U$–$J_H/U$ phase diagram, mapping the superconducting domes ($U<U_c$) of $\tau_z$-triplet (orange) and $\tau_0$-triplet (green) pairing against the altermagnetic (AM, blue) and ferromagnetic (FM, gray) ordered phases ($U \geq U_c$). The red dashed line marks the point where $\chi^{\text{RPA}}_{\text{AM}}(\Gamma)$ equals $\chi^{\text{RPA}}_{\text{FM}}(\Gamma)$.
}
\label{fig4}
\end{figure}

To establish the generality of our conclusions, we map the phase diagram as a function of $U$ and $J_H$. First, at a fixed $U = 0.99U_c$, the ratio $\tilde{\Delta}_{yy}/\tilde{\Delta}_{xx}$ is computed versus $J_H/U$ [Fig.~\ref{fig4}(a)]. This ratio is negative (close to $-1$) in the dominant AM fluctuation regime, signifying $\tau_z$-triplet pairing, and becomes positive (near $+1$) in the dominant FM fluctuation regime, consistent with $\tau_0$-triplet pairing. Near the AM-FM boundary (dashed gray line), $\tilde{\Delta}_{yy}/\tilde{\Delta}_{xx}$ deviates from these ideal values due to strong mixing between the $\tau_z$ and $\tau_0$ channels, yet it still exhibits a sharp sign change across the transition.

The full phase diagram in the $U$–$J_H/U$ plane is presented in Fig.~\ref{fig4}(b). 
For $U \geq U_c$, the divergence rate of $\chi_{\text{AM}}^{\text{RPA}}(\Gamma)$ and $\chi_{\text{FM}}^{\text{RPA}}(\Gamma)$ as $U\to U_c$ determines whether AM (blue) or FM (gray) order develops, with a critical ratio $J_H/U = 0.133$ (gray dashed line). 
Below $U_c$, the dominant fluctuation is identified by directly comparing $\chi_{\text{AM}}^{\text{RPA}}(\Gamma)$ and $\chi_{\text{FM}}^{\text{RPA}}(\Gamma)$, with the boundary marked in red. The corresponding superconducting states, namely, $\tau_z$-triplet (orange) and $\tau_0$-triplet (green), are indicated. These results indicates a direct correspondence between the dominant magnetic fluctuation and the symmetry of the resulting superconducting state. Additional computational details are provided in Sec.~D of SM.

\begin{figure}[t]
\centering
\includegraphics[width=0.48\textwidth]{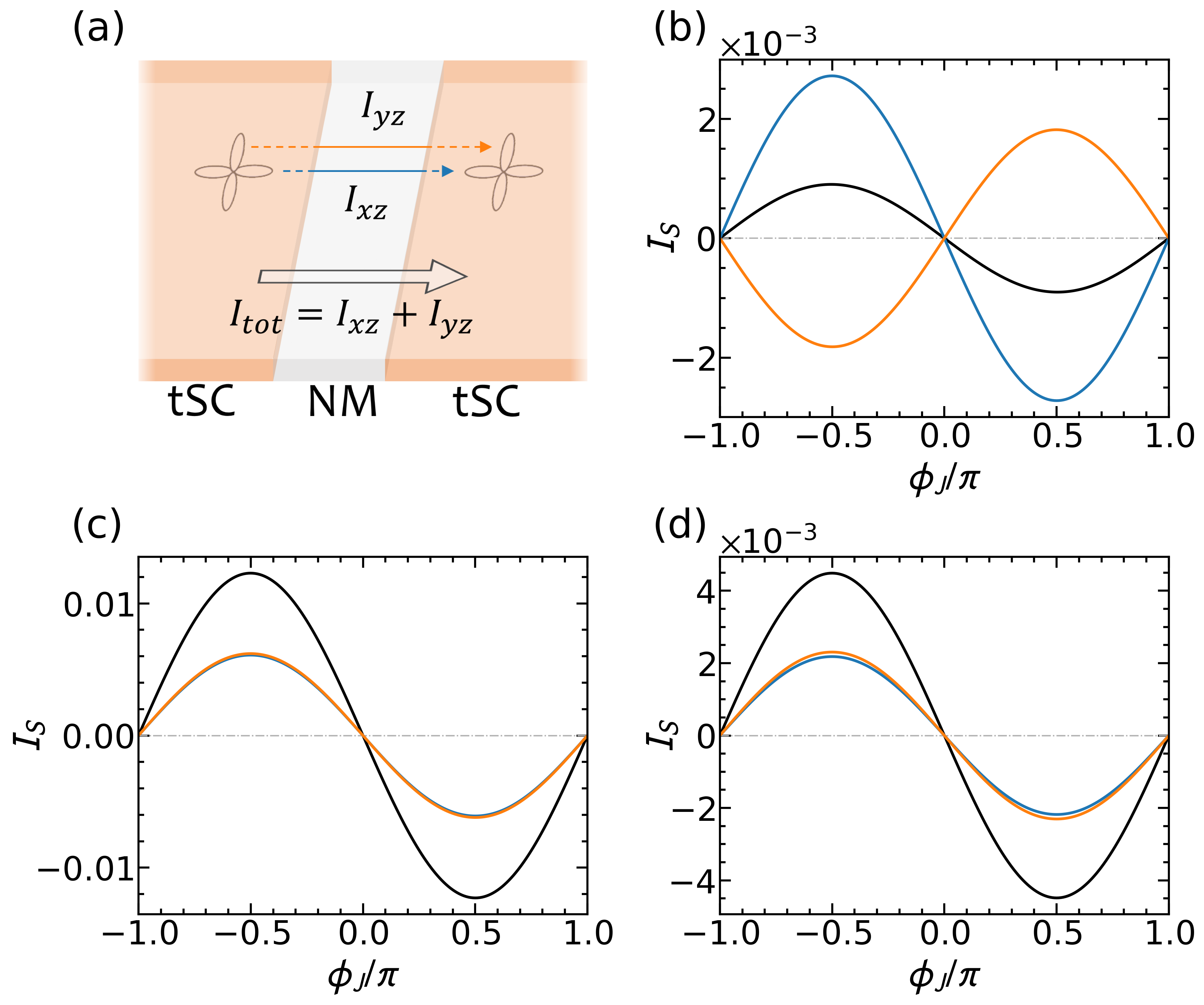}
\caption{Possible experimental signatures.
(a) Schematic of a Josephson junction between two spin-triplet superconductors (tSCs) separated by a normal metal (NM). The size $(L_{\text{SC}},L_{\text{NM}})=(200,30)$.
(b-d) Current-phase relations $I(\phi_J)$ for different configurations: (b) $\tau_z$-$\tau_0$, (c) $\tau_0$-$\tau_0$, and (d) $\tau_z$-$\tau_z$. The pronounced suppression of the supercurrent uniquely in the $\tau_z$-$\tau_0$ junction (b) provides a clear fingerprint of the $\tau_z$-triplet state, distinguishing it from the conventional $\tau_0$-triplet.
}
\label{fig5}
\end{figure}

\paragraph{{\color{blue}Triplet-triplet Josephson junction}}
As established above, the $\tau_z$-triplet state represents a distinct class of spin-triplet superconductivity characterized by its unique orbital structure. To demonstrate its fundamental difference from conventional $\tau_0$-triplet pairing and provide experimental signatures, we investigate triplet-triplet Josephson junctions [Fig.~\ref{fig5}(a)]. We consider three representative scenarios: (1) $\tau_z$-$\tau_0$ junctions, (2) $\tau_0$-$\tau_0$ junctions, and (3) $\tau_z$-$\tau_z$ junctions. In our planar junction setup, the normal-metal (N) region is designed with zero inter-orbital hybridization to suppress orbital-flip scattering. Due to current conservation across the junction, the total supercurrent in the normal region decomposes into orbital-resolved components,
\begin{align}
I_{\text{tot}}(\phi_J) = I_{d_{xz}}(\phi_J) + I_{d_{yz}}(\phi_J),
\end{align}
where $I_{d_{xz}}$ and $I_{d_{yz}}$ denote the supercurrent contributions through the $d_{xz}$ and $d_{yz}$ orbitals, respectively, and $\phi_J$ denotes the Josephson phase difference. We first analyze an idealized limit where inter-orbital hybridization is also absent within the superconducting electrodes. In this case, the orbital-resolved lowest order currents are given by $I_{d_{xz}}(\phi_J) = I_{d_{yz}}(\phi_J) = \Delta_{L}\Delta_{R} \cos \phi_J$ for cases (2-3), while $I_{d_{xz}}(\phi_J) = -I_{d_{yz}}(\phi_J) = \Delta_{L}\Delta_{R} \cos \phi_J$ for case (1). Crucially, only in $\tau_z$-$\tau_0$ junctions does the internal $\pi$-phase difference characteristic of the $\tau_z$-triplet state lead to exact cancellation of intra-orbital supercurrents, resulting in a vanishing total Josephson current.

We next perform realistic simulations using the full BdG Hamiltonian in Eq.~\eqref{eq-bdg-hamsc}, where inter-orbital hybridization is restored in the superconducting regions. Following the continuity equation [see Sec.~E of SM], we compute the supercurrent flowing across the junction~\cite{Asano2001PRB,Sakurai17PRB,SBZhang20PRB}. As shown in Figs.~\ref{fig5}(b-d), our calculations confirm that although the exact cancellation in $\tau_z$-$\tau_0$ junctions is lifted due to orbital mixing in  superconductors, a pronounced suppression of the total supercurrent remains as a distinctive signature. This suppression effect occurs uniquely in $\tau_z$-$\tau_0$ junctions, providing a clear experimental fingerprint for identifying the $\tau_z$-triplet state.
However, strong inter-orbital hybridization or electron doping may weaken this effect.

\paragraph{{\color{blue}Conclusion}}
We note that while an isolated $\tau_z$-triplet channel might support chiral $p_x \pm i p_y$ superconductivity, our full RPA calculations take all intra-orbital, inter-orbital, and pair-hopping interactions into account, and may reveal a different ground state. For $J_H/U=0.06$, we find a nematic spin-triplet superconducting state, where the dominant $\tau_z$-triplet component cooperates with longer-range subdominant pairings, precluding the formation of a chiral state [see Sec.~F of SM].

In summary, we have established altermagnetic fluctuations as a distinct mechanism for spin-triplet superconductivity. The inversion symmetry breaking that interchanges the two orbitals induces momentum-orbital locking, which suppresses the inter-orbital singlet channel. Crucially, a subdominant fluctuation acts as an internal Josephson coupling and mediates a unique $\tau_z$-triplet state—fundamentally an inter-orbital spin-triplet superconductor that stands in sharp contrast to the conventional $\tau_0$-triplet from ferromagnetic fluctuations. Our findings thus establish a new fluctuation-pairing correspondence, expanding the landscape of spin-fluctuation-mediated superconductivity and highlighting altermagnetism as a promising route to novel triplet superconductors with non-trivial orbital structure.

\paragraph{{\color{blue}Acknowledge}}
We thank Zhan Wang, Fan Yang, Jian Kang, and Gabriel Aeppli for helpful discussions.

\vspace{2pt}
\noindent
\textbf{C.~Lu and C.~Li contributed equally to this work.}

\bibliography{references}

\end{document}